\documentclass[preprint,onecolumn,nofootinbib]{revtex4}
\usepackage[colorlinks=true,linkcolor=blue,urlcolor=blue,filecolor=black,citecolor=red,pdfstartview=FitV,pdftitle={},pdfsubject={},pdfkeywords={},pdfpagemode=None,bookmarksopen=true]{hyperref}
\usepackage{graphicx}
\usepackage{amsmath}
\usepackage{amsfonts}
\usepackage{mathrsfs} 
\usepackage{amssymb,ulem}
\usepackage{color}%
\usepackage{tikz}
\usepackage{dcolumn}
\usepackage{listings}
\usepackage{xcolor}

\begin{document}
\title{
  Screened volume law of Holographic Entanglement Entropy in Holographic Spontaneous Vectorization model
}
\author{Chong-Ye Chen $^{1}$}
\email{cychen@stu2022.jnu.eud.cn}
\author{Guang-Zai Ye $^{1}$}
\email{photony@stu2022.jnu.edu.cn}
\author{Peng Liu $^{1}$}
\email{phylp@email.jnu.edu.cn}
\thanks{corresponding author}
\affiliation{
  $^1$ Department of Physics and Siyuan Laboratory, Jinan University, Guangzhou 510632, China
}

\begin{abstract}

  We present a holographic study of spontaneous vectorization in the background of an isotropic asymptotically AdS black brane. By extending spontaneous scalarization to vector fields, we demonstrate how the effective mass of the vector field drives tachyonic instability, leading to a transition from the AdS-RN phase to a vectorized phase. Thermodynamic analysis reveals the critical temperature $ T_c $ and coupling $ \alpha_c $ for this transition, with the vectorized phase exhibiting lower free energy. A central discovery is the emergence of a ``screened volume law'' in the holographic entanglement entropy (HEE), a novel phenomenon where the entanglement entropy scales with the subregion size via a screened entropy density distinct from thermal entropy. This arises from a geometric constraint tied to the vanishing of the Christoffel symbol $ \Gamma^z{}_{xx} $, which defines an effective boundary outside the horizons. Unlike conventional ``entanglement shadows'' in black hole systems, this surface acts as a boundary for minimal surfaces in a translationally invariant geometry. This screening effect suggests the inability of entanglement measure to fully probe the Hilbert space of this thermal system. Additionally, the HEE in the vectorized phase displays non-monotonic temperature dependence. These results establish spontaneous vectorization as a mechanism for generating novel entanglement structures in holographic systems, with implications for quantum information and critical phenomena in strongly coupled systems.  

\end{abstract}

\maketitle

\tableofcontents

\section{Introduction}\label{sec:introduction}

Holography has emerged as a powerful framework for probing strongly coupled systems, rooted in the celebrated AdS/CFT correspondence. This duality provides a profound link between weakly coupled gravitational theories in asymptotically AdS spacetimes and strongly coupled CFTs defined on the boundary. Over the past two decades, this paradigm has expanded beyond its string-theoretic origins, yielding insights into diverse areas such as condensed matter physics (AdS/CMT) and quantum chromodynamics (AdS/QCD) \cite{Donos:2012js,Donos:2014oha,Ling:2015epa,Kiritsis:2015oxa,Baggioli:2014roa,Arefeva:2018hyo,Domokos:2007kt}. A fertile direction has been the study of spontaneous symmetry breaking—a universal mechanism underpinning phenomena like superconductivity and superfluidity \cite{Hartnoll:2008kx,Gubser:2008px,Hartnoll:2008vx,Horowitz:2010gk,Cai:2012nm,Cai:2014ija,Hartnoll:2016apf}. Holographic models of symmetry breaking, often mediated by the condensation of matter fields, have provided a controlled setting to analyze nonlinear dynamics and phase transitions in strongly coupled systems. Recently, spontaneous scalarization—a mechanism where scalar field condensation is triggered by non-minimal couplings to gauge field has emerged as a novel avenue for holographic model-building \cite{Guo:2021zed,Chen:2023iws,Chen:2023eru}.

Quantum information measures have become essential tools for analyzing the structure and dynamics of strongly coupled quantum systems through the lens of holography. A cornerstone of these tools is the holographic entanglement entropy (HEE), which provides a geometric prescription of entanglement entropy (EE) to quantify quantum correlations in boundary CFTs via the area of minimal surfaces in the bulk spacetime \cite{Ryu:2006bv}. This prescription not only deepens our understanding of quantum systems but also provides tools for analyzing phenomena such as thermal phase transitions and quantum phase transitions \cite{Cai:2013oma,Yang:2023wuw,Ling:2015dma,Ling:2016wyr,Dudal:2018ztm,Baggioli:2020cld}. HEE serves as a sensitive probe of phase transitions, capturing the change of entanglement as systems evolve between distinct phases. However, the application of HEE faces limitations in certain backgrounds. For instance, when probing the bulk geometry of BTZ black hole, regions near the event horizon—dubbed ``entanglement shadow''—remain inaccessible to minimal surfaces anchored at the boundary \cite{Hubeny:2013gta,Freivogel:2014lja,Balasubramanian:2014sra}. This complicates the reconstruction of bulk spacetime from boundary entanglement data, challenging the assertion of the encoding between geometry and entanglement \cite{Swingle:2009bg,VanRaamsdonk:2010pw}. To circumvent these limitations, recent advances have proposed generalized entanglement measures such as entwinement \cite{Balasubramanian:2014sra,Gerbershagen:2021gvc,Gerbershagen:2024hkt,Craps:2022pke}. 

In thermal systems, the entanglement entropy (EE) typically obeys a volume law, scaling proportionally to the thermal entropy for sufficiently large subregions. However, in many-body localized (MBL) systems, the entanglement entropy grows with system size but remains subthermal, falling short of the values expected for a fully thermalized state, as noted in studies \cite{Dumitrescu:2017sto,Iyer:2013}. This subthermal behavior is also observed in other contexts, including systems with dynamical scar states, dipole-conserving Hamiltonians, and entanglement phase transitions \cite{Pai:2019rfq,Sala:2019zru,Ippoliti:2021iju}. A unifying feature of these systems is their inability to fully explore the available Hilbert space, or incomplete thermalization of the system.

In our work, we extend the phenomenon of spontaneous scalarization—previously studied in scalar fields—to vector fields within an isotropic AdS black brane spacetime, employing holographic techniques to investigate the phenomenon of spontaneous vectorization, as inspired by previous work \cite{Oliveira:2020dru}. Unlike scalar fields, vector fields introduce additional complexities due to their coupling to the metric and the potential for extra instabilities like ghost and gradient instabilities \cite{Himmetoglu:2009qi,Clough:2022ygm}. We derive the perturbation equation for the vector field and compute its quasinormal modes (QNMs) to identify critical tempereatures and coupling constants for the onset of vectorization. The tachyonic instability, signaled by a negative effective mass squared, drives the system to a new vectorized phase.

Next, our investigation centers on the thermodynamic and entanglement properties of this spontaneous vectorization model. The phase transition, governed by a critical temperature and coupling constant, reveals a thermodynamically preferred vectorized phase, as determined through free energy. A striking finding is the behavior of the HEE, where we identify a novel ``screened volume law.'' This phenomenon bears a resemblance to the subthermal volume law observed in MBL and related systems. However, the mechanism in our model is distinct: the entanglement entropy scales linearly with the subregion width, yet the entropy density is reduced compared to the thermal case. We attribute this suppression to a zero point in a specific component of the Christoffel symbol, which creates a ``shadow region'' outside the horizon inaccessible to minimal surfaces. 

This work is organized as follows. In Section \ref{sec:emv_setup}, we introduce the holographic setup for spontaneous vectorization, including the action and equations of motion. Section \ref{sec:perbanalysis} presents a detailed perturbation analysis, identifying the critical points for the vectorization. In Section \ref{sec:thermalandgeometrical}, we examine the thermodynamical properties of the vectorized phase based on numerical data, including free energy and entropy density, and discuss the emergent zero point of Christoffel symbol. Section \ref{sec:hee} explores the entanglement in the context of spontaneous vectorization, focusing on the screened volume law of HEE and entanglement of vectorized phase. Finally, we conclude with a discussion of the broader implications of our findings and potential future directions in Section \ref{sec:discuss}.

\section{Setup of the Holographic Spontaneous Vectorization Model}\label{sec:emv_setup}

In this section, we introduce a holographic spontaneous vectorization by considering the following action \cite{Oliveira:2020dru},
\begin{equation}\label{action}
  \mathcal{S}=\frac1{4\pi}\int d^4x\sqrt{-g}\left[\frac R4 - \frac{\Lambda}{2} -\frac{f(|B|^2)}4F^{\mu\nu}F_{\mu\nu}-\frac14G^{\mu\nu}G_{\mu\nu}^*\right],
\end{equation}
where $R$ is the Ricci scalar, $F_{\mu\nu}=\partial_\mu A_\nu-\partial_\nu A_\mu$ is the field strength of the gauge field $A_\mu$ and $G_{\mu\nu}=\partial_\mu B_\nu-\partial_\nu B_\mu$ correspond to the complex vector field $B_\mu$. The coupling function $f(|B|^2)$, which depends on the norm $|B|^2=B_\mu B^\mu$, ensures consistency with the Einstein-Maxwell theory when $ B_\mu = 0 $, provided $f(0)=1$. Varying the action with respect to $ B_\mu $ yields a Proca-like equation, 
\begin{equation}\label{veceom}
  \nabla_\mu G^{\mu\nu}=\frac12\frac{df}{d|B|^2}F^2B^\nu ,
\end{equation}
where $ F^2 \equiv F^{\mu\nu} F_{\mu\nu} $. This introduces an effective mass term,
\begin{equation}\label{effpotential}
  \mu_{eff}^2=\frac{1}{2}\frac{df}{d|B|^2}F^2,
\end{equation}
despite the absence of an explicit mass for $ B_\mu $ in the original action. In the probe limit, where backreaction of $ B_\mu $ is neglected, a perturbative expansion $ f(|B|^2) \approx 1 + \alpha |B|^2 + \cdots $ leads to $ \mu_{\text{eff}}^2 \propto \alpha F^2 $. For $ \alpha \neq 0 $, the effective mass-squared can become negative in an electric vacuum (e.g. RN-AdS balck brane) background where $ F^2 < 0 $, triggering a tachyonic instability. This signals a spontaneous vectorization—a phase transition where the vector field condenses spontaneously. To satisfy the condition $ \mu_{\text{eff}}^2 < 0 $ for $ F^2 < 0 $, the coupling must obey $ df/d|B|^2 > 0 $. A minimal alternative fulfilling this requirement is the exponential coupling 
\begin{equation}\label{coupling}
  f(|B|^2)=e^{\alpha|B|^2},
\end{equation}
where coupling constant $\alpha>0$. 

The fully nonlinear behavior of vector field $B_\mu$ into this gravitational system is described by the following equations,
\begin{equation}\label{eoms}
  \begin{aligned}
    R_{\mu\nu}-\frac12g_{\mu\nu}R+\Lambda g_{\mu\nu} &= 2T_{\mu\nu} \\
    \nabla_\mu(fF^{\mu\nu}) &= 0
  \end{aligned}
\end{equation}
where the energy-momentum tensor $T_{\mu\nu}$ is given by
\begin{equation}\label{emt}
  \begin{aligned}
    T_{\mu\nu}=& f(|B|^2){\left(F_\mu\right.}^\alpha F_{\nu\alpha}-\frac14g_{\mu\nu}F^{\alpha\beta}F_{\alpha\beta})+\frac12\left(G_\mu{}^\alpha G_{\nu\alpha}^*+G_\mu^{*\alpha}G_{\nu\alpha}-\frac12g_{\mu\nu}G^{\mu\nu}G_{\mu\nu}^*\right) \\
    &+\frac14\frac{df}{d|B|^2}F^{\alpha\beta}F_{\alpha\beta}\left(B_\mu B_\nu^*+B_\mu^*B_\nu\right) .
    \end{aligned}
\end{equation}

To study spontaneous vectorization in an asymptotically AdS black brane, we adopt a homogeneous ansatz compatible with the no-hair theorem constraints for vector fields \cite{Oliveira:2020dru}. The metric and gauge field are given by
\begin{equation}\label{ansatz1}
  \begin{aligned}
    &ds^2=\frac1{z^2}\left(-p(z)(1-z)U(z)dt^2+\frac1{p(z)(1-z)U(z)}dz^2+V(z)dx^2+V(z)dy^2\right)\\&
    A_\nu dx^\nu=\mu(1-z)A_t(z)dt
  \end{aligned}
\end{equation}
where $p(z)=1+z+z^2-\mu^2 z^3$ and $\mu$ is the chemical potential, the coordinates are $\{t,z,x,y\}$ and the radial coordinate $ z \in [0,1] $ spans from the AdS boundary ($ z=0 $) to the horizon ($ z=1 $). This ansatz is isotropic in the $x$ and $y$ directions. The vector field ansatz mirrors the gauge field structure,
\begin{equation}\label{ansatz2}
  B_\mu dx^\mu=\mu (1-z) B_t(z)dt ,
\end{equation}
ensuring the isotropy of background geometry. The Reissner-Nordström (RN) AdS black brane solution corresponds to $ B_t = 0 $, $ U = V = A_t = 1 $, with the Dirichlet condition $ U(1) = 1 $ fixing horizon geometry for the convenience of parameter analysis. The dimensionless temperature is $ T = (\mu^2 - 3)/(4\pi \mu) $, where $ \mu $ is set as the energy scale and unit of physical quantities. The dimensionless parameter space is thus characterized by the coupling constant $ \alpha $ and $ T $.

\section{Perturbation Analysis of Spontaneous Vectorization and Dynamical Instability}\label{sec:perbanalysis}

\begin{figure}
  \centering
  \includegraphics[height=0.4\textwidth]{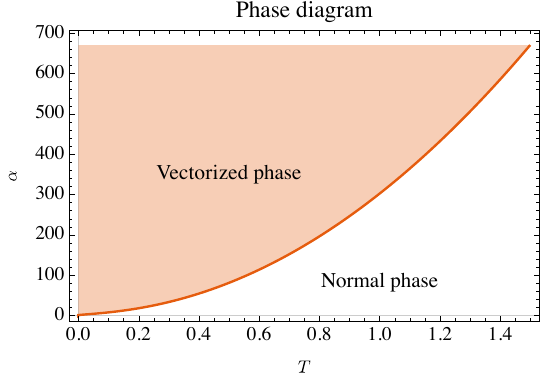}
  \caption{The phase diagram of the spontaneous vectorization in the asymptotic AdS black brane. The critical coupling constant $\alpha_c$ for each critical temperature $T_c$ is obtained by solving the eigenvalue problem using $80$ points in Chebyshev-Gauss-Lobatto grid. The results are shown as the orange line between vectorized phase and AdS-RN phase. The numerical results remain stable as the density of grid points increases.}
  \label{fig:phasediagram}
\end{figure}

In this section, we analyze the linear perturbation of the spontaneous vectorization. The onset of this instability is determined by the effective mass, which induces a tachyonic instability in the probe limit. The phase boundary separating the normal phase (RN-AdS black brane) from the vectorized phase is identified by solving the eigenvalue problem derived from the linearized equation of motion for the vector field $ B_t(z) $,
\begin{equation}\label{perbphaseeom}
  \alpha \mu^2 z^2  B_t(z) + \left(-1 - z(1 + z) + \mu^2 z^3 \right) \left(2  B_t'(z) + (-1 + z) B_t''(z)\right)=0
\end{equation}
where $\prime$ represents the derivative of the function with respect to $z$. The boundary condition $ B_t(0) = 0 $, reflecting regularity at the boundary, motivates the redefinition $ B_t(z) = z \hat{B}_t(z) $. This allows the system to be a stable eigenvalue problem for the coupling constant $\alpha$. The phase boundary $ \{\tilde{T}_c, \alpha_c\} $ is depicted in Fig. \ref{fig:phasediagram}.  
Crossing the boundary from AdS-RN phase side to the vectorized phase side, the system will be unstable under the small perturbation of the vector field, and then the vector field will grow exponentially and the system transitions to the vectorized phase, resulting in the spontaneous vectorization. 

While phase diagram provide a partial understanding of spontaneous vectorization, a complete characterization requires analyzing dynamical instabilities through quasinormal mode (QNM). To study this, we adopt ingoing Eddington-Finkelstein coordinates for the metric and fields,
\begin{equation}\label{ansatz3}
  \begin{aligned}
    ds^2         &= \frac1{z^2}\left(-p(z)(1-z)U(z)dv^2-2dtdz+V(z)dx^2+V(z)dy^2\right)\\
    A_\nu dx^\nu &= \mu(1-z)A_t(z)dv \\
    B_\nu dx^\nu &= B_{0,\epsilon} dv + B_{0,\epsilon}/((1-z)p(z)U(z)) dz
  \end{aligned}
\end{equation}
where $dv = dt - dz/((1-z)p(z)U(z))$ and $B_{0,\epsilon}=\mu(1-z)B_t(z)+\epsilon b_t(z)e^{-i \omega v}$.
The gauge freedom eliminates the $ z $-component of $ A_\mu $, but not for $B_\mu$, due to explicit quadratic terms in the action. The perturbation $ b_t(z) = z \hat{b}_t(z) $ ensures regularity at the AdS boundary, while the $ e^{-i\omega v} $ factor enforces ingoing boundary conditions at the horizon. Substituting this ansatz into the equations of motion yields a linearized eigenvalue equation for $ \hat{b}_t(z) $, with $ \omega^* = \omega/\mu $ as the dimensionless frequency. We solve this eigenvalue problem numerically, focusing on fundamental QNM frequencies for AdS-RN black branes below the critical temperature $ \tilde{T}_c $. 
\begin{figure}
  \centering
  \includegraphics[height=0.4\textwidth]{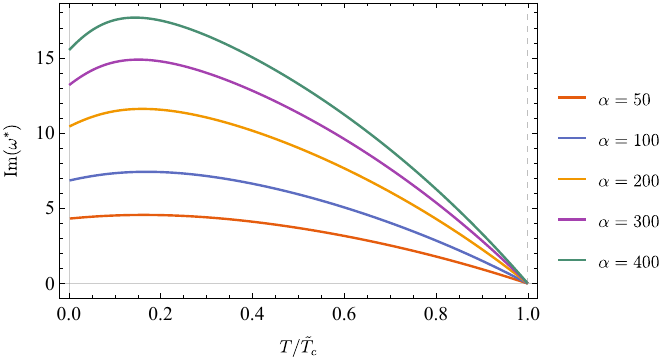}
  \caption{The imaginary part of the fundamental modes, i.e. the leading behavior of damping or growing of the vector perturbation. We also apply the Chebyshev-Gauss-Lobatto grid to solve these frequencies. All of these fundamental modes are numerically stable under the variation of the grid points. Observe that, the imaginary part of the fundamental modes are all positive, which means the system is dynamically unstable under the small perturbation of the vector field, resulting in the exponential growth of the vector field. This is consistent with the spontaneous vectorization mechanism.}
  \label{fig:SVQNMRN}
\end{figure}
As shown in Fig. \ref{fig:SVQNMRN}, the imaginary parts of these frequencies are positive, indicating exponential growth after small perturbations in vector field. This dynamical instability aligns with the spontaneous vectorization mechanism that temperatures below the critical point ($ T < T_c $) destabilize the normal phase, triggering a transition to the vectorized phase. The growth rate of vector field (proportional to $ \text{Im}(\omega^*) $) increases monotonically with the coupling constant $ \alpha $.

\section{Thermodynamical and Geometrical Properties of Asymptotically AdS Vectorized Black Brane}\label{sec:thermalandgeometrical}

\subsection{Thermodynamical properties and criterion of phase transition}
\label{sec:thermodynamic}

\begin{figure}
  \centering
  \includegraphics[height=0.27\textwidth]{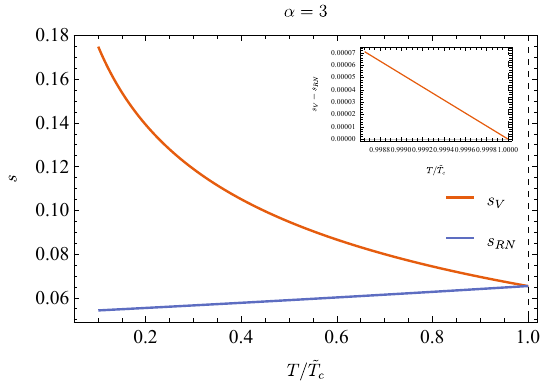}
  \includegraphics[height=0.265\textwidth]{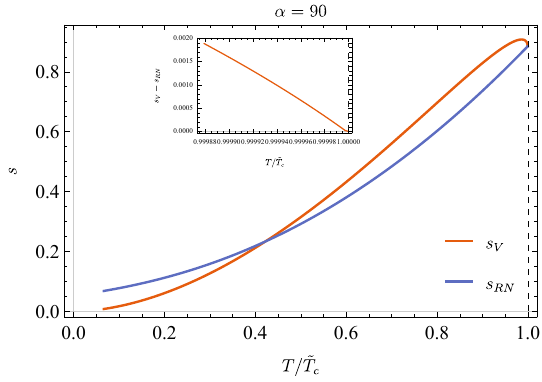}
  \includegraphics[height=0.27\textwidth]{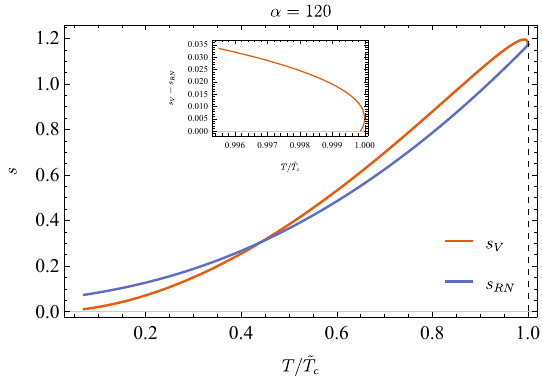}
  \includegraphics[height=0.27\textwidth]{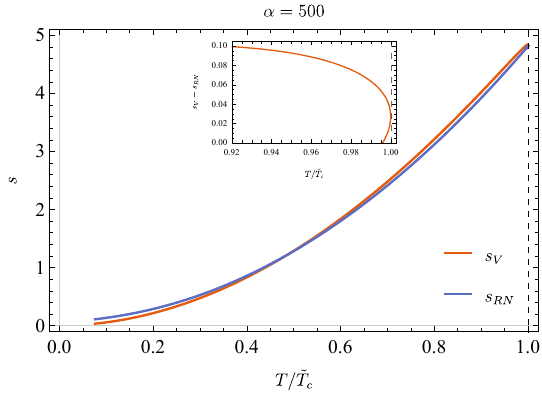}
  \caption{The entropy density as the function of temperature ratio $T/T_c$ for both normal phase and vectorized phase. The inset plots show the entropy density difference between the two phases near the critical temperature. The entropy density of the vectorized phase is always higher than the normal phase for small coupling constant $\alpha=3$, while for larger coupling constant $\alpha=90, 120, 500$, the entropy density of the vectorized phase is higher than the normal phase near the critical temperature.}
  \label{fig:sda3a90a120a500}
\end{figure}

To investigate the thermodynamic stability of the vectorized and normal phases, we analyze the entropy density and free energy. The dimensionless entropy density is defined as
\begin{equation}
  s = \frac{\hat s}{\mu} = \frac{V(1)}{2\pi \mu}.
\end{equation}
Fig. \ref{fig:sda3a90a120a500} illustrates the entropy density for both phases across varying coupling constants $\alpha$. For small coupling constant ($\alpha=3$), the entropy density of the vectorized phase exceeds that of the normal phase and remains finite as $T \to 0$. At intermediate coupling ($\alpha = 90$), the entropy density of the vectorized phase initially rises but decreases at low temperatures. For larger coupling constants ($\alpha = 120, 500$), the vectorized phase persists above the perturbative critical temperature and exhibits higher entropy density near this critical point, indicating thermodynamic preference over the normal phase. Consequently, the entropy criterion suggests a critical temperature higher than the perturbative prediction (see the inset plots in Fig. \ref{fig:sda3a90a120a500}). However, at sufficiently low temperatures, the entropy density of the vectorized phase becomes smaller than that of the normal phase, implying a possible transition to the normal state. This competition between phases highlights the necessity of free energy analysis to complete the equilibrium phase structure, as entropy alone does not universally dictate stability across all temperature regimes.

\begin{figure}
  \centering
  \includegraphics[height=0.27\textwidth]{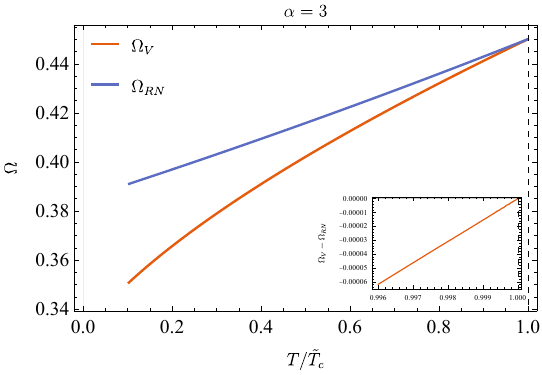}
  \includegraphics[height=0.27\textwidth]{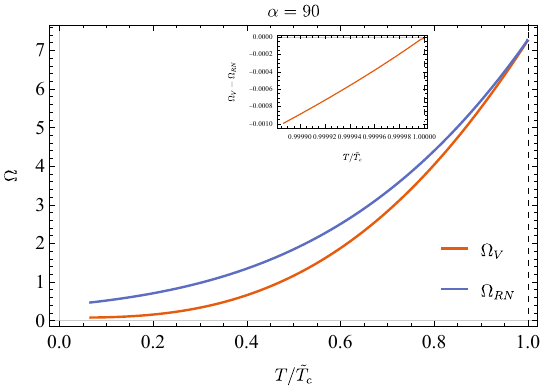}
  \includegraphics[height=0.27\textwidth]{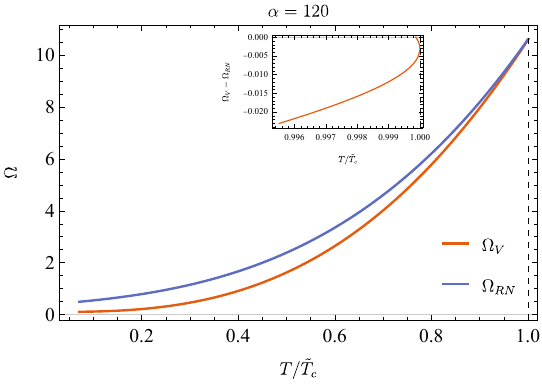}
  \includegraphics[height=0.27\textwidth]{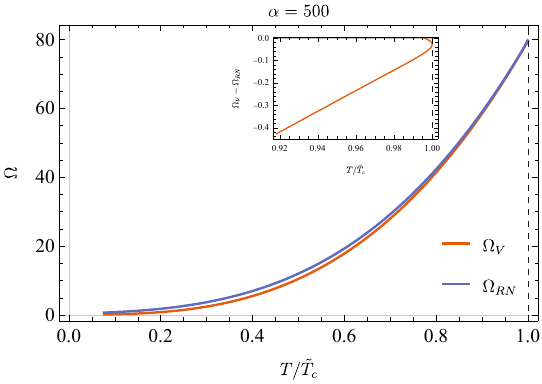}
  \caption{The free energy as the function of temperature ratio $T/T_c$ for both normal phase and vectorized phase. The inset plots show the free energy difference between the two phases near the critical temperature. The free energy of the vectorized phase is always lower than the normal phase, indicating the thermodynamic preference of the vectorized phase.}
  \label{fig:FEa3a90a120a500}
\end{figure}

To compute the free energy of the system, we analyze the near-boundary expansion of the metric component $ g_{tt} $
\begin{equation}
  g_{tt}  \sim  -\frac{1}{z^2} - \frac{U'(0)}{z} - \frac{U''(0)}{2}+\left(1+\mu  -\frac{1}{6} U^{(3)}(0)\right) z + \cdots,
\end{equation}
where $ U(0) = 1 $. The coefficient of linear term in $z$, which corresponds to $-2\hat{M}$, defines the dimensionless black brane mass $ M = \hat{M}/\mu^2 = -\frac{1}{2}\left( (\mu + 1) - \frac{1}{6} U'''(0) \right) $. The free energy is then determined thermodynamically as
\begin{equation}
  F = M - T s. 
\end{equation}
Numerical results for free energy, comparing the normal and vectorized phases across varying coupling constant, are presented in Fig. \ref{fig:FEa3a90a120a500}. 

Crucially, the vectorized phase exhibits a lower free energy than the normal phase for all $ \alpha $, indicating a thermodynamic preference for vectorization below the critical temperature. This persists as $ T \to 0 $, confirming the stability of the vectorized phase in the zero-temperature limit. The absence of any free energy crossings at low temperatures further suggests that no additional phase transitions occur in this regime. These results align with the expectation that spontaneous vectorization, driven by the coupling $ \alpha $, represents the equilibrium configuration in the low-temperature regime of this holographic system.

The critical temperature of the phase transition, $ T_c $, is determined by analyzing the free energy and entropy density. Specifically, $ T_c $ corresponds to the maximum temperature at which the non-trivial vectorized phase with lower free energy. The comparison of the critical temperature based on the perturbation analysis and the free energy criterion is presented in Table \ref{tab:Tccompare}. 
\begin{table}[h]
  \centering
  \begin{tabular}{|c|c|c|c|c|c|}
  \hline
     & $\alpha=3$  & $\alpha=90$ & $\alpha=120$ & $\alpha=500$  \\ \hline
  perturbation analysis    & $\tilde T_c=0.02875$    & $\tilde T_c=0.5292$ & $\tilde T_c=0.6177$ & $\tilde T_c=1.2926$    \\ \hline
  free energy criterion  & $T_c=0.02875$    & $T_c=0.5292$ & $T_c=0.6178$ & $T_c=1.2978$    \\ \hline
  \end{tabular}
  \caption{Critical temperatures $ T_c $ computed using perturbation analysis and the free energy criterion for varying coupling constants $\alpha$.}
  \label{tab:Tccompare}
\end{table}

The results exhibit strong agreement for coupling constants below $\alpha \approx 103.8$.\footnote{Due to the rapid growth of the vector field near the phase transition, this behavior complicates the precise determination of the critical $\alpha$ value, as the vector field dynamics become increasingly sensitive to small perturbations in the parameter space} For larger coupling constants ($\alpha > 103.8$), there are a difference between two method in the critical temperature. This discrepancy arises from the existence of two branches of the vectorized phase above the critical temperature (as shwon in the inset plots in both Fig. \ref{fig:sda3a90a120a500} and Fig. \ref{fig:FEa3a90a120a500}). And the phase transition is a zeroth order phase transition, which means the free energy will jump from the normal phase to the vectorized phase at the critical temperature. 

\subsection{The charge and vector charge of the vectorized phase}

\begin{figure}
  \centering
  \includegraphics[height=0.35\textwidth]{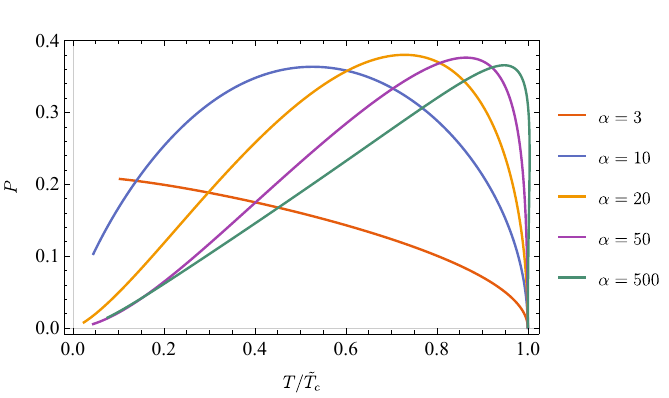}
  \caption{Vector charge $ P $ in the vectorized phase as a function of $ T/\tilde T_c $ for different coupling constants $ \alpha $.}
  \label{fig:Pa3toa500}
\end{figure}

The near-AdS boundary expansions of the Maxwell field and vector field are given by
\begin{equation}
  \begin{aligned}
    (1-z)A_t(z) &\sim 1 + \left(A_t'(0) - 1\right)z + \cdots, \\  
    (1-z)B_t(z) &\sim B_t'(0) \, z + \cdots,
  \end{aligned}
\end{equation}
where the coefficients of the linear $ z $-terms are the electric charge $ Q $ and vector charge $ P $, respectively. The presence of a non-zero $ P $ serves as the signal of spontaneous vectorization. Fig. \ref{fig:Pa3toa500} illustrates the temperature-dependent behavior of $ P $ across varying coupling $ \alpha $.  
For small coupling constant, the vector charge $P$ increases monotonically with the decreasing temperature. For larger coupling constant, however, the vector charge $P$ initially increases with the decreasing temperature and then decreases at low temperature, ultimately showing a vanishing trend as $T \to 0$.

\begin{figure}
  \centering
  \includegraphics[height=0.25\textwidth]{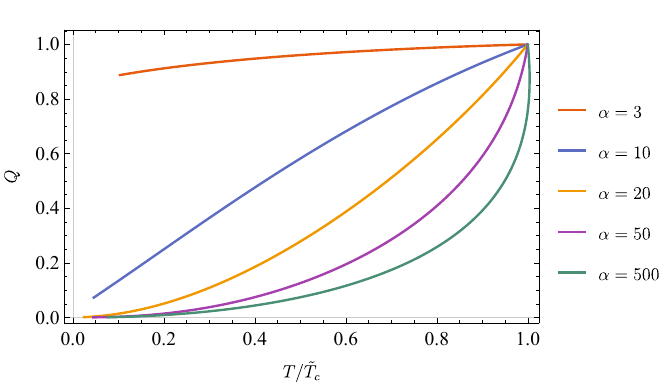}
  \includegraphics[height=0.25\textwidth]{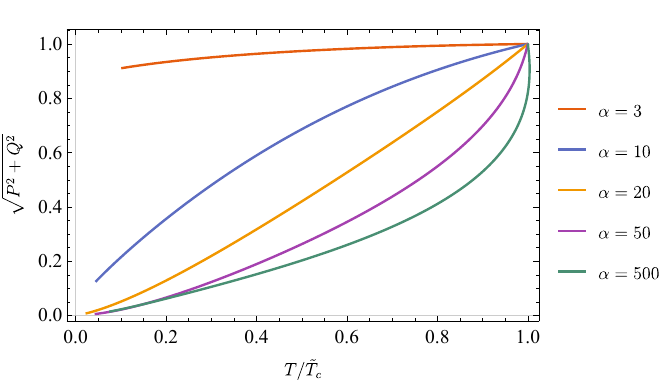}
  \caption{Electric charge $ Q $ (left) and combined charge $ \sqrt{Q^2 + P^2} $ (right) in the vectorized phase as functions of $ T/\tilde T_c $ for varying $ \alpha $.}
  \label{fig:QCCa3toa500}
\end{figure}

Given the structural similarity between the Maxwell and vector fields, the interplay between $ Q $ and $ P $ needs systematic study. A particularly useful lens is the combined charge $ \sqrt{Q^2 + P^2} $ \cite{Ye:2024pyy}, which, along with $ Q $, is plotted in Fig. \ref{fig:QCCa3toa500}. Below $ T_c $, $ Q $ decreases monotonically with decreasing temperature for all $ \alpha $, with larger couplings constant amplifying this suppression. The combined charge follows a similar trend, decreasing steadily at lower $ T $. Notably, for $ \alpha > 10 $, both $ Q $, $ P $, and $ \sqrt{Q^2 + P^2} $ vanish asymptotically as $ T \to 0 $. 

Crucially, the combined charge in the vectorized phase remains strictly smaller than the electric charge $ Q = 1 $ of the normal phase (dimensionless unity charge of RN-AdS black brane is fixed by the selected ansatz), underscoring the universal suppression of charge carriers by non-minimal coupling. This suppression intensifies with $ \alpha $, revealing a robust interplay between the coupling and the charge.  

\subsection{The Analysis for the Existence of Zero Point of Christoffel Symbol}\label{sec:chrzxx}

The Christoffel symbols associated with the metric-compatible connection $\nabla_a$ in the coordinate system $\{t,z,x,y\}$ are defined as
\begin{equation}
\Gamma^c{}_{ab}=\frac12g^{cd}(\partial_ag_{bd}+\partial_bg_{ad}-\partial_dg_{ab}),
\end{equation}
where $g_{ab}=g_{\mu\nu}(dx^\mu)_a(dx^\nu)_b$. For the specific case of $\Gamma^z{}_{xx}$, applying the definition and noting the symmetry properties yields
\begin{equation}
  \Gamma^z{}_{xx}=-\frac12 z^2 g^{zz} \partial_z g_{xx},
\end{equation}
with $g^{zz}=  z^2 (1-z) \left(-\mu ^2 z^3+z^2+z+1\right)U(z) $ and $g_{xx} = V(z)/z^2 $. Notably, $g^{zz}$ vanishes at the horizon $z=1$, ensuring $\Gamma^z{}_{xx}$ also vanishes there — a generic feature of black brane geometries. While $g_{xx}$ remains finite everywhere, its $z$-derivative $\partial_z g_{xx}$ may vanish at specific point, indicating non-monotonicity of $g_{xx}$. 

\begin{figure}
  \centering
  \includegraphics[height=0.4\textwidth]{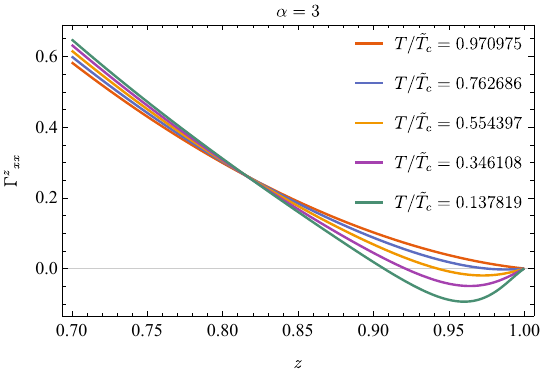}
  \caption{Christoffel symbol component $\Gamma^z{}_{xx}$ near the horizon. The zero point of $\Gamma^z{}_{xx}$ indicates the effective boundary that minimal surface can not penetrate.}
  \label{fig:chrzxxdemoa3}
\end{figure}

The existence of roots for $\partial_z g_{xx}$ is constrained by boundary conditions derived from the EoMs. Near the horizon $z=1$, asymptotic analysis gives
\begin{equation}
  g_{xx}'(1)=-\frac{2 g_{xx}(1) \left(3- \left(A_t(1)^2+B_t(1)^2 \right)\mu^2\right)}{3-\mu^2},
\end{equation}
where $\mu^2 < 3$ and $g_{xx}(1) > 0$. For the normal phase, $A_t(1)=1$ and $B_t(1)=0$, leading to $g_{xx}'(1)<0$. In contrast, vectorized phases at lower temperatures modified $A_t(1)$ and $B_t(1)$, potentially satisfying $\left(A_t(1)^2 + B_t(1)^2\right)\mu^2 > 3$, which flips the sign of $g_{xx}'(1)$ to positive. At the AdS boundary ($z \to 0$), $g_{xx} \sim 1/z^2$ implies $g_{xx}'(0) < 0$. If $g_{xx}'(1) > 0$, continuity necessitates an intermediate root for $\partial_z g_{xx}$, rendering $g_{xx}$ non-monotonic. As a result, $\Gamma^z{}_{xx}$ also vanishes at this point, as seen in Fig. \ref{fig:chrzxxdemoa3}.
Based on this properties, some geodesic curves may not penetrate the region near the horizon, such as the minimal surface. This leads to a novel phenomenon in the entanglement of this system, which we will explore in the following section.

\section{Holographic Entanglement Entropy in spontaneous vectorization}\label{sec:hee}

\subsection{Setup of Holographic Entanglement Entropy in asymptotically AdS black brane}
\label{sec:hee_setup}

Entanglement entropy serves as a foundational quantity in quantum mechanics to characterize the entanglement between a subsystem $A$ and its complement $B$ across all quantum systems, from simple few-body arrangements to complex quantum field theories. It is defined as the von Neumann entropy of the reduced density matrix $ \rho_A $, obtained by tracing out the degrees of freedom in $ B $:  
\begin{equation}
  S_\text{A}=-\text{Tr}(\rho_A \log \rho_A).
\end{equation}

Within the framework of the AdS/CFT correspondence, the holographic entanglement entropy (HEE) provides a geometric realization of this entropy in terms of bulk spacetime dynamics. For a state in the boundary CFT dual to the AdS geometry, the HEE is computed by the Ryu-Takayanagi formula \cite{Ryu:2006bv},
\begin{equation}
  S_\text{A}=\frac{\text{Area}(\gamma_A)}{4G_N}.
\end{equation}
where $ \gamma_A $ denotes the minimal surface in the bulk spacetime that is anchored on the boundary $ \partial A $ of the subregion. 

The area of the minimal surface in the bulk spacetime can be systematically computed as follows. Consider a minimal surface with translational symmetry along the spatial boundary direction $ y $. Parametrizing the minimal surface by its profile $ z(x) $ in the radial direction $ z $ and transverse direction $ x $, and thus the Lagrangian density reads,
\begin{equation}\label{xlagrangian}
  \hat L \left(z(x),z'(x),x\right) = \sqrt{g_{yy}(z(x))\left(g_{xx}(z(x))+ g_{zz}(z(x))z'(x)^2\right)}.
\end{equation}
where $z(x)$ is the generalized coordinate, $z'(x) \equiv dz/dx$ is the generalized velocity and $x$ is treated as the curve parameter. Since the background spacetime is translation invariant along the $x$ direction, the hamiltonian density
\begin{equation}
  \hat H = \frac{\partial \hat L}{\partial z'(x)}z'(x)-\hat L
\end{equation} 
is conserved along the minimal surface. This conservation law implies
\begin{equation}\label{constantH}
  \hat H(z(x),z'(x),x)=\hat H(z^*,z^*{}'(x^*),x^*)=\text{const}.
\end{equation}
where $ (z^*, x^*) $ denotes the turning point (also the nearest point to the horizon) of the surface, defined by $z^*{}'(x)=0$. By symmetry, we fixes $ x^* = 0 $ without loss of generality. Substituting $ z'(x^*) = 0 $ simplifies the conserved Hamiltonian to
\begin{equation}
  \hat H(z(x),z'(x),x)=-g_{xx}(z^*),
\end{equation}
which constraints $z'(x)$ via
\begin{equation}\label{conszpx}
  z'(x)=\pm\frac{\sqrt{g_{xx}(z(x))^3-g_{xx}(z^*)^2g_{xx}(z(x))}}{g_{xx}(z^*)\sqrt{g_{zz}(z(x))}}.
\end{equation}
Here, the $\pm$ sign corresponds to the left ($ x<0 $) and right ($x>0$) branches of the minimal surface. The entanglement entropy $ S_{\text{HEE}} $ and the boundary subregion width $ w $ are then determined by integrating over the radial coordinate $ z $
\begin{equation}
  \begin{aligned}
  S_\text{HEE}=2\int_0^{z^*} \frac{\hat L}{z'(x)}dz,\\ 
  w=2\int_0^{z^*} \frac{1}{z'(x)}dz.
  \end{aligned}
\end{equation}
where the factor of $2$ accounts for the two symmetric halves of the surface. Near the AdS boundary, the integrals of $S_{\text{HEE}}$ become divergent. To obtain finite, physically meaningful results, these divergences require regularization, typically achieved by introducing a cutoff or subtracting the divergent terms. These integrals depend parametrically on the turning point $ z^* $, which itself is fixed by the boundary condition specifying the width $ w $ of the subregion $ A $.

\subsection{Analysis for effective boundary of minimal surfaces}

In this subsection, we will discuss a general property of the HEE in vectorized phase, focusing the geometric constraints imposed by the zero point of the Christoffel symbol component $\Gamma^z{}_{xx}$. We consider the minimal surface anchored on a boundary interval, parameterized by a coordinate $\theta \in [0,\pi]$, with the reparameterized Lagrangian, 
\begin{equation}
  \tilde{L}(z(\theta),x(\theta),z'(\theta),x'(\theta),\theta)=\sqrt{g_{yy}(z(\theta ))\left(g_{xx}(z(\theta )) x'(\theta)^2+g_{zz}(z(\theta )) z'(\theta)^2\right)},
\end{equation}
where primes denote derivatives with respect to $\theta$. The $x$-translation symmetry ensures a conserved canonical momentum,
\begin{equation}
  \frac{\partial \tilde{L}}{\partial x'(\theta)}=\frac{g_{xx}(z(\theta)) g_{yy}(z(\theta)) x'(\theta)}{\sqrt{g_{yy}(z(\theta)) \left(g_{xx}(z(\theta)) x'(\theta)^2 + g_{zz}(z(\theta)) z'(\theta)^2\right)}}=\text{const}.
\end{equation}
This conservation law implies $x'(\theta)$ maintains a fixed sign along the minimal surface. 

\begin{figure}
  \centering
  \includegraphics[height=0.4\textwidth]{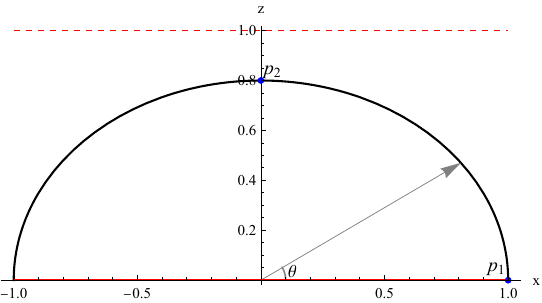}
  \caption{HEE demonstration}
  \label{fig:heedemo}
\end{figure}

To fix the parametrization freedom, we impose the parameter constraint,
\begin{equation}
  \sin (\theta ) x(\theta )-\cos (\theta ) z(\theta )=0.
\end{equation}
which relates $\theta$ to the $(x,z)$ coordinates (Fig. \ref{fig:heedemo}).
AdS geometry requires $x'(0) = 0$ at the anchor point $p_1$ ($\theta = 0$). The minimal surface exhibits mirror symmetry across the $x=0$ plane, ensuring smoothness at the equatorial turning point $p_2$ ($\theta = \pi/2$). Here, the derivatives satisfy
\begin{equation}\label{consonehalfpi}
    z'(\pi/2)=0, \quad x'(\pi/2)=-z^*.
\end{equation}
These conditions enforce 
\begin{equation}\label{consxptheta}
  x'(\theta)\leqslant 0 
\end{equation}
globally. Combining the constraint Eq. \eqref{conszpx}, we can obtain the following relation, 
\begin{equation}
  \begin{aligned}
  z'(\theta)&\geqslant 0 \quad\quad (\theta \in [0,\pi/2]) \\
  z'(\theta)&\leqslant 0 \quad\quad (\theta \in [\pi/2,\pi]) \\
  \end{aligned}
\end{equation}

Let us examine the geodesic equations governing the minimal surface, expressed in terms of $x(\theta)$ and $z(\theta)$,
\begin{equation}
  \begin{aligned}
    x''(\theta )+2  \Gamma
     ^x{}_{zx} z'(\theta ) x'(\theta )&=0\\
    z''(\theta )+ \Gamma
     ^z{}_{xx}x'(\theta )^2+ \Gamma ^z{}_{zz}z'(\theta
     )^2&=0  
  \end{aligned}
\end{equation}
At the turning point $p_2$, the boundary conditions \eqref{consonehalfpi} reduce the $z$-equation to
\begin{equation}\label{eq:zequation}
  z''(\pi/2)+ \Gamma^z{}_{xx}x'(\pi/2 )^2 =0.
\end{equation}
For standard cases, $\Gamma^z{}_{xx}(z)\geqslant 0 $ through the bulk, and vanishes only at the horizon, as illustrated by the red line in Fig. \ref{fig:chrzxxdemoa3}. For large boundary subregions, the minimal surface penetrates deeply, approaching $z^* \to 1$. Since $\Gamma^z{}_{xx}(z^*) \to 0$ near the horizon, $z''(\pi/2) \to 0$, and $z''(\theta) \leqslant 0$ everywhere due to $\Gamma^z{}_{xx} \geqslant 0$. This forces $z'(\theta)$ to decrease monotonically to zero at $p_2$, rendering the surface asymptotically tangent to the horizon. The dominant area contribution arises from this near-horizon region, ensuring the HEE scales as
\begin{equation}
  S_{\text{HEE}} \propto s \cdot w,
\end{equation}
consistent with expectations from entanglement entropy in QFT.

Now consider a non-standard scenario where $\Gamma^z{}_{xx}(z)$ has an additional zero at $z=z_b<1$, as depicted by the green curve Fig. \ref{fig:chrzxxdemoa3}. This zero introduces an ``effective boundary'' in the spacetime, altering the behavior of the minimal surface. Below this effective boundary, where $\Gamma^z_{xx} \geqslant 0$, the condition $z''(\pi/2) \leqslant 0$ holds, and as the surface approaches this point $z_b$, $\Gamma^z_{xx} \to 0$, yielding $z''(\pi/2) \to 0$. This mirrors the property of horizon, implying that the minimal surface becomes tangent to this effective boundary rather than crossing it. To see why the surface cannot penetrate this boundary, assume it crosses into a region where $\Gamma^z_{xx} < 0$. If the surface crossed into a region where $\Gamma^z_{xx} < 0$, the equation at the turning point would yield 
\begin{equation}
  z''(\pi/2) = -\Gamma^z_{xx} x'(\pi/2)^2 > 0,
\end{equation}
since $x'(\pi/2)^2 > 0$. A positive $z''(\pi/2)$ implies that $z'(\theta)$ would increase near the turning point $p_2$, remaining positive as $\theta \to \pi/2$. This contradicts the boundary condition $z'(\pi/2) = 0$, which ensures the smoothness of the minimal surfaces. Thus, crossing the effective boundary is geometrically inconsistent, constraining the surface to remain above or tangent to it. 

Consequently, this effective boundary, though not a true horizon or singularity, imposes a horizon-like role and screens the minimal surface. The HEE then follows a modified scaling, termed a ``screened volume law''
\begin{equation}
  S_\text{HEE}\propto \bar s \cdot w,
\end{equation}
where $\bar{s}$ denotes a ``screened entropy density'' tied to the geometric properties of this effective boundary rather than the thermal entropy of the horizon.

\subsection{The numerical analysis of screened volume law of HEE}

\begin{figure}
  \centering
  \includegraphics[height=0.26\textwidth]{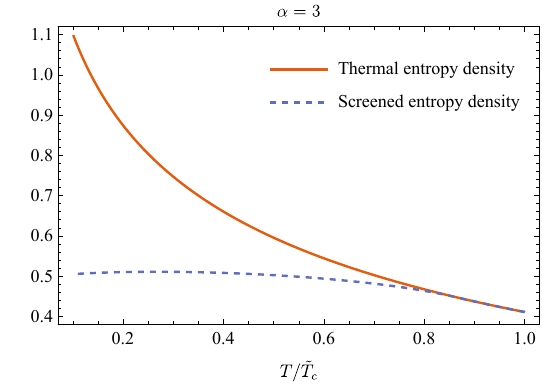}
  \includegraphics[height=0.255\textwidth]{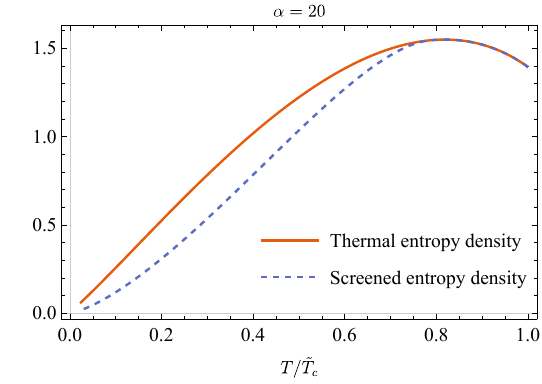}
  \caption{Thermal entropy density and screened entropy density for selected cases. To clearly illustrate the screened volume law, we consider coupling constant $\alpha=3$ and $\alpha=20$ as representative examples. }
  \label{fig:psda3a20}
\end{figure}

The thermal entropy density $s$ determines the full size of the phase space, while the screened entropy density $\bar s$ characterizes the effective phase space available to the boundary states. In Fig. \ref{fig:psda3a20}, we compare the thermal entropy density and the screened entropy density. For small values of the coupling constant, $\bar s$ exhibits non-monotonic behavior with respect to temperature, in contrast to the thermal entropy density, which increases monotonically. For a larger coupling constant, such as $\alpha = 20$, the factor remains consistently smaller than the thermal entropy density, with their difference initially increasing and then decreasing as the temperature is reduced. 

\begin{figure}
  \centering
  \includegraphics[height=0.26\textwidth]{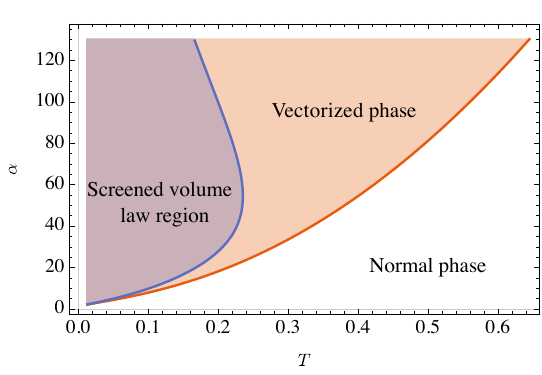}
  \includegraphics[height=0.245\textwidth]{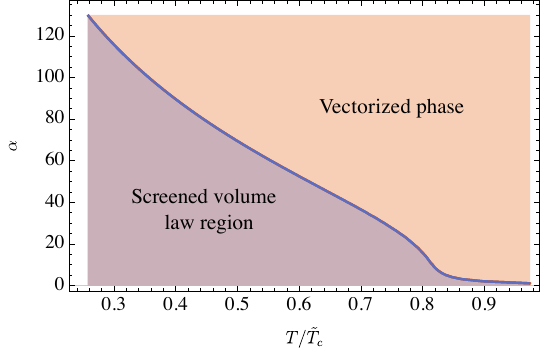}
  \caption{The existence domain boundary of the screened volume law as a function of the coupling constant $\alpha$, shown for both the dimensionless temperature $T$ and the temperature ratio $T/T_c$.}
  \label{fig:exidomain}
\end{figure}

In Fig. \ref{fig:exidomain}, we illustrate the domain where the screening effect exists, plotted against the coupling constant $\alpha$ for both the dimensionless temperature $T$ and the temperature ratio $T/T_c$. The blue region denotes the existence domain of the screening effect, embedded within the orange region corresponding to the vectorized phase. We define $T_b$ as the temperature at which the effective boundary emerges in the bulk, noting that $T_b < T_c$. As $\alpha$ increases, the existence domain initially expands and subsequently contracts, differing from the monotonic increase of $T_c$. The ratio $T_b / T_c$, always less than $1$, decreases monotonically with increasing $\alpha$, indicating that larger coupling constants hinder the formation of the effective boundary in the bulk spacetime. 

\begin{figure}
  \centering
  \includegraphics[height=0.36\textwidth]{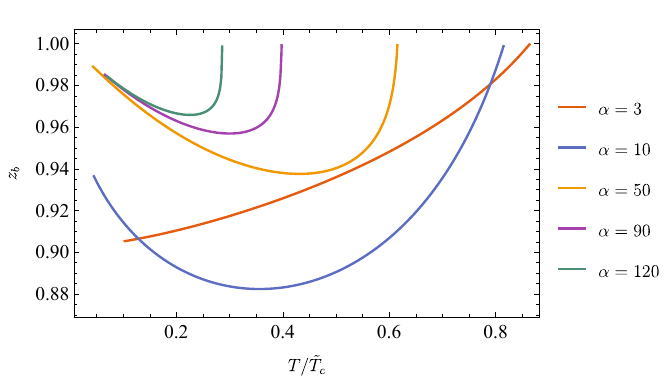}
  \caption{Radial coordinate $z_b$ of the effective boundary as a function of temperature ratio $T/T_c$ for couplings $\alpha=3,10,50,90,120$.}
  \label{fig:Tzba3a120}
\end{figure}

We investigate the dependence of effective boundary $z_b$, on temperature for various values of the coupling constant $\alpha$, as shown in Fig. \ref{fig:Tzba3a120}. For a small coupling constant, such as $\alpha=3$, the effective boundary shifts toward the AdS boundary as the temperature decreases, leading to a amplification of screening effect. In contrast, for a larger coupling constant, such as $\alpha=10$, the behavior of $z_b(T)$ exhibits non-monotonicity: the effective boundary initially shifts outward, then inward, with decreasing temperature. For an even larger coupling constant, such as $\alpha=120$, the effective boundary moves outward more rapidly at higher temperatures before gradually shifting inward toward the horizon as the temperature continues to decrease. 
Despite this effective boundary is more related to the horizon geometry, the shrinking of the shadow region in low temperature for larger coupling constant is consistent with the vanishing charge and combined charge of the vectorized phase shown in Fig. \ref{fig:QCCa3toa500}, indicating the connection between the charge degrees of freedom and the screening mechanism. When the coupling constant is samll ($\alpha = 3$ or $\alpha = 10$), increasing $\alpha$ leads to a more significant screening effect. However, as the coupling constant becomes larger ($\alpha = 50$), the degrees of freedom related to the vector field and Maxwell field are suppressed, weakening the screening effect. 

\subsection{The numerical result of HEE across the entire the vectorized phase}
\label{HEE_res}

\begin{figure}
  \centering
  \includegraphics[height=0.26\textwidth]{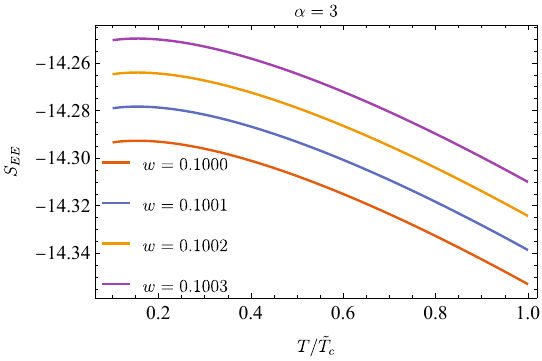}
  \includegraphics[height=0.26\textwidth]{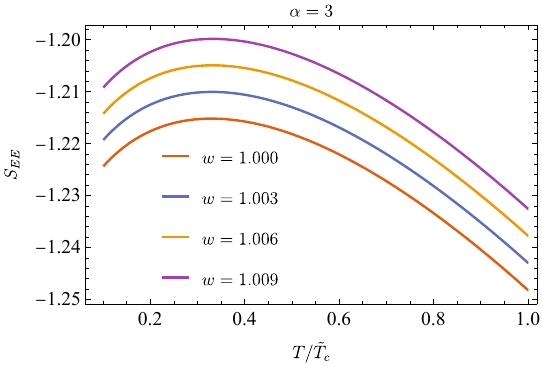}
  \includegraphics[height=0.26\textwidth]{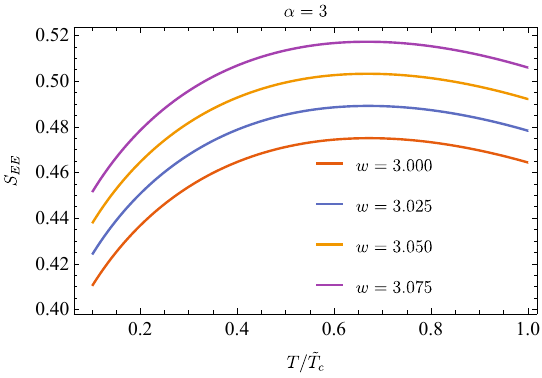}
  \includegraphics[height=0.26\textwidth]{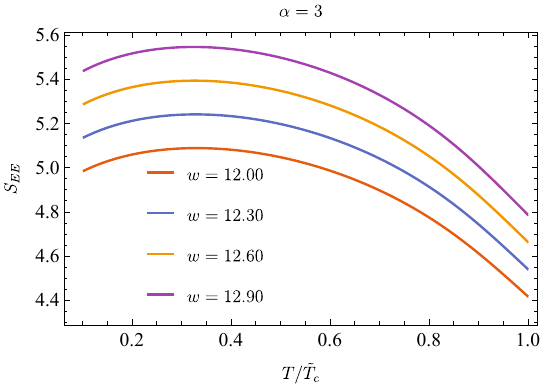}
  \caption{The HEE as the function of temperature ratio $T/T_c$ corresponding to various subregion width, in small coupling constant ($\alpha=3$).}
  \label{fig:heea3}
\end{figure}

Due to the screening effect, the computation of the HEE is restricted to the region $[0, z_b]$. The HEE exhibits distinct temperature-dependent behavior across different coupling strengths and subsystem sizes. For the case of $\alpha=3$, as illustrated in Fig. \ref{fig:heea3}, the HEE of a small subregion with width $w=0.14$ increases nearly linearly with decreasing temperature, though it exhibits slight non-monotonic deviations. As the subregion size grows, incorporating a greater number of degrees of freedom, this non-monotonic behavior becomes more pronounced. Contrary to initial expectations, the HEE for a large subregion ($w=9$) does not display a monotonic increase with decreasing temperature, as might be anticipated from the entropy density. Instead, it follows a non-monotonic trajectory that closely corresponds to the screened entropy density, as depicted in Fig. \ref{fig:psda3a20}.

\begin{figure}
  \centering
  \includegraphics[height=0.26\textwidth]{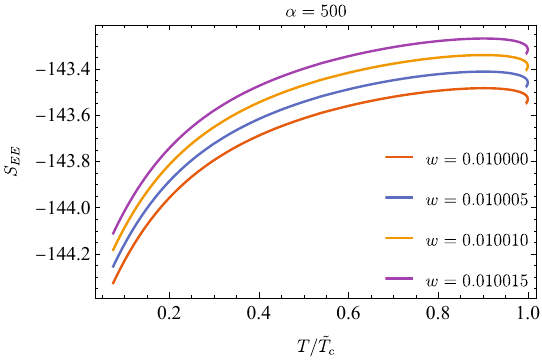}
  \includegraphics[height=0.26\textwidth]{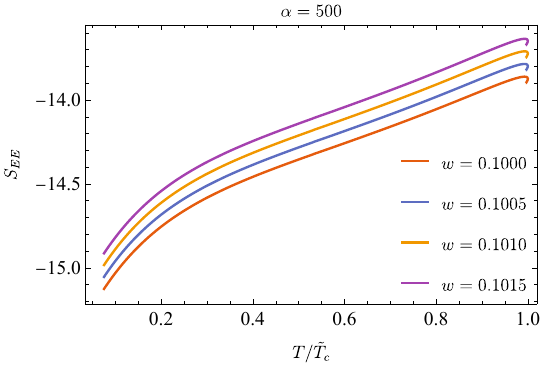}
  \includegraphics[height=0.26\textwidth]{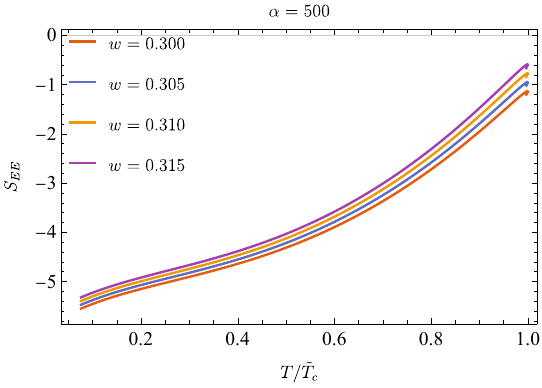}
  \includegraphics[height=0.26\textwidth]{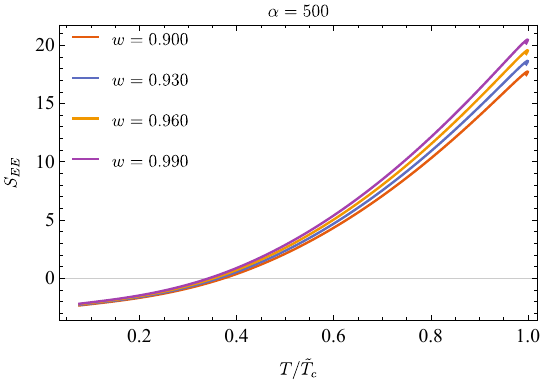}
  \caption{The HEE as the function of temperature ratio $T/T_c$ corresponding to various subregion width, in large coupling constant ($\alpha=500$).}
  \label{fig:heea500}
\end{figure}

In contrast, when the coupling constant is increased to $\alpha=500$, as shown in Fig. \ref{fig:heea500}, the HEE for a small subregion ($w=0.1$) exhibits marked non-monotonic behavior with decreasing temperature, characterized by a convex functional form. As the subregion size increases, the temperature dependence of the HEE transitions from convex to concave. For larger subregions, such as $w=0.9$, the HEE demonstrates a more significant decrease with decreasing temperature, aligning with the behavior of the entropy density, as observed in the bottom right panel of Fig. \ref{fig:sda3a90a120a500}. Moreover, at lower temperatures, the HEE values for different subregion sizes converge, consistent with the decreasing entropy density of the vectorized phase. Analysis of the background geometry in this case ($\alpha=500$) reveals the absence of an effective boundary, which also indicate that HEE obeys the standard volume law.

\section{Discussion}\label{sec:discuss}

In this work, we generalize spontaneous scalarization to vector fields within AdS/CFT, leveraging the AdS/CFT duality to study phase transitions in strongly coupled systems. The mechanism of spontaneous vectorization arises from a tachyonic instability triggered when the effective mass-squared of the vector field becomes negative, inducing a transition from an AdS-RN black brane (normal phase) to a vectorized phase. Critical parameters for this transition—the temperature $T_c$ and coupling constant $\alpha_c$—were determined through linear perturbation and quasinormal mode analyses. 

Thermodynamic consistency is confirmed by comparing free energies: the vectorized phase becomes energetically favorable below $T_c$, with entropy density exhibiting non-monotonic temperature dependence at larger couplings. This aligns with the interpretation of vectorization as a thermodynamically driven symmetry-breaking process stabilized at lower temperatures. Notably, as the coupling constant exceeds a critical value $\alpha \approx 103.8$, the phase transition becomes zeroth order.

The vectorized phase displays a screened volume law for holographic entanglement entropy, characterized by a suppressed entropy density below the thermal entropy density. This distinctive scaling originates from a geometric shadow region located outside the event horizon, delineated by a codimension-$2$ surface termed the effective boundary, where the metric component \( \Gamma^z_{xx} \) vanishes. Unlike horizons or singularities, this boundary imposes a condition on minimal surfaces, a feature stemming from the non-minimal coupling between the vector and Maxwell fields without the presence of singularities. Generalization of entanglement entropy measures—entwinement—cannot probe this shadow region, as extremal surfaces fail to wind this infinite planar boundary. Consequently, the entanglement wedge also remains restricted to the exterior, leading to effective screening of boundary state entanglement despite the full thermalization of this system. This screening mechanism depends critically on both the vector field degrees of freedom and the strength of the non-minimal coupling.

As the key parameter in this holographic model, the coupling constant $ \alpha $ exerts a nuanced and multifaceted influence over the system behavior, defying a simplistic classification as merely a ``weak'' or ``strong'' coupling parameter. When $ \alpha $ is small, it amplifies deviations in the entropy density between the normal and vectorized phases and the screening effect—an unexpected outcome tied to the structure of the coupling function. This function, expressed as
\begin{equation}
  \begin{aligned}
    \exp\left( \alpha |B|^2 \right)  &= \exp\left( - \frac{\alpha \mu^2 (1-z) z^2 B_t(z)^2}{\left(1 + z + z^2 - \mu^2 z^3\right) U(z)} \right) ,
  \end{aligned}
\end{equation}
with $\left(1 + z + z^2 - \mu^2 z^3\right)>0$, reveals how the couplings are modulated: a smaller $ \alpha $ can enlarge certain contributions, leading to a raise in the effective coupling strength in specific contexts. Conversely, a larger $ \alpha $ enhances the tachyonic instability by modifying the effective mass, which in turn expands the parameter space where the vectorized phase persists. This intricate interplay also manifests in the suppression of the screening effect and a reduction of charge-related degrees of freedom at low temperatures. 

Though structurally analogous to couplings in spontaneous scalarization, the observed phenomena reveal a richer functional role for $ \alpha $ in holography. Extensions of this framework to systems with broken translational symmetry or lattice-structured vector fields could elucidate novel quantum phase transitions and entanglement dynamics. Connections between the screening effect and quantum information measures—such as entanglement wedge cross sections or computational complexity—may further deepen our understanding. Additionally, the emergent shadow region hints at connections to non-trivial horizon geometry, motivating investigations into Lyapunov exponents and butterfly velocities, which encode scrambling dynamics in black hole physics. Such explorations may unify insights into entanglement, chaos, and criticality in strongly correlated systems.  

\section*{Acknowledgments}

Peng Liu would like to thank Yun-Ha Zha and Yi-Er Liu for their kind encouragement during this work. This work is supported by the Natural Science Foundation of China under Grant No. 12475054 and Guangdong Basic and Applied Basic Research Foundation No. 2025A1515012063.

\appendix

\section{The technical details of the numerical method}\label{app:numerical}

In this model, we solve the planar AdS black brane background from the highly non-linear second-order ordinary differential equations (ODEs) that dictate the behavior of the metric functions and matter fields. Our approach integrates spectral methods with iterative techniques. Specifically, we discretize the radial coordinate $ z $ using Chebyshev collocation points and solve the resulting ODEs employing the Newton-Raphson method. For the numerical solutions, we adopt a grid of $300$ collocation points and apply a stretching transformation defined as
\begin{equation*}
  z \rightarrow \tanh^{-1}(\tanh(3) z).
\end{equation*}
This transformation clusters grid points near the horizon while maintaining analytic tractability at the AdS boundary. And thus this enhances convergence efficiency, particularly in the near-horizon region at low temperatures. The iterative procedure was terminated when the residual error fell below $ 10^{-6} $, with numerical precision set at single \texttt{MachinePrecision} throughout the computation.  

To determine the quasi-normal modes (QNMs), we cast the problem into an eigenvalue framework using Chebyshev collocation. We configure the computation with $70$ collocation points and validate the results by cross-checking with outcomes obtained using $80$ points. In this investigation, the QNMs are computed within the RN-AdS background, where the near-horizon geometry remains regular and analytically manageable. For numerical black hole backgrounds exhibiting singular geometries, a stretching method would be essential to maintain convergence.

\end{document}